\documentstyle[prd,aps,amsfonts]{revtex} 
 
\newcommand{\jump}[1]{{\left[ #1 \right]}} 
\newcommand{\avrg}[1]{{\left\langle #1 \right\rangle}} 
\newcommand{\dm}{n} 
\newcommand{\dd}{{\rm d}} 
\newcommand{\norm}{n} 
\newcommand{\kapp}{\kappa^2} 
\newcommand{\kap}{\kappa}

\newcommand{\eps}{\varepsilon} 
\newcommand{\emt}{\eps} 
\newcommand{\ivm}{\eta} 
\newcommand{\tn}{\overline{\cal T}_{_\infty}} 
\newcommand{\bulk}{\overline{\cal W}} 
\newcommand{\force}{\bar{f}} 
\newcommand{\tfree}{\omega} 
\newcommand{\order}[1]{{\mathcal{O}}\left(#1\right)} 
\newcommand{\brR}{\overline{R}} 
\newcommand{\bkR}{{\cal R}} 
\newcommand{\brF}{{\cal F}} 
\newcommand{\brG}{\overline{G}} 
\newcommand{\brdel}{\overline{\cal D}} 
\newcommand{\den}{\hat{\rho}} 
\newcommand{\prs}{\hat{p}} 
\newcommand{\s}{\sigma} 
\newcommand{\al}{\alpha} 
\newcommand{\be}{\beta} 
 
\newcommand{\bom}{\bbox{\omega}}

\title{Einstein equations for an asymmetric brane-world} 
\author{Richard A. Battye${}^{1}$, Brandon Carter${}^{2}$,  
Andrew Mennim${}^{1}$ and Jean--Philippe Uzan${}^{3}$\vskip0.25cm} 
\address{${}^{1}$ Department of Applied Mathematics and Theoretical 
Physics, Centre for Mathematical Sciences \\ University of Cambridge, 
Wilberforce Road, Cambridge, CB3 0WA, UK \\ 
${}^{2}$ D\'{e}partment d'Astrophysique Relativiste et de Cosmologie, 
UMR 8629 du CNRS, \\ 
 Observatoire de Paris, 92195 Meudon Cedex, France \\ 
${}^{3}$ Laboratoire de Physique Th\'{e}orique, UMR 8627 du CNRS, \\  
Universit\'{e} de Paris XI, B\^atiment 210, 91405 Orsay 
Cedex, France} 
\date{10 May 2001} 
 
\begin{document} 
 
\maketitle 
 
\begin{abstract} 
We consider a brane-world of co-dimension one without the reflection
symmetry that is commonly imposed between the two sides of the
brane. Using the coordinate-free formalism of the Gauss-Codacci
equations, we derive the effective Einstein equations by relating the
local curvature to the matter on the brane in the case when its bare
tension is much larger than the localized matter, and hence show that
Einstein gravity is a natural consequence of such models in the weak
field limit. We find agreement with the recently derived cosmological
case, which can be solved exactly, and point out that such models can be
realized naturally in the case where there is a minimally coupled form
field in the bulk.
\end{abstract} 
 
\section{Introduction} 
\label{sec-intro} 
 
Much recent work has focused on the idea that our 4 dimensional universe
might be a 3-brane in a higher dimensional
spacetime~\cite{ADD,RS1}. Although the initial motivation was to use the
size of the extra dimensions to set the gravitational and electroweak
scales equal in the higher dimensional space, hence ``solving'' the
hierarchy problem, possibly the most striking result is that, under
certain circumstances, higher dimensional gravity can appear from the
point of view of an observed localized on the brane to be similar to
ordinary Einstein gravity, with (i) the Newtonian inverse square law in
the non-relativistic regime~\cite{RS2} above a given length scale and
(ii) a cosmology compatible with the standard
picture~\cite{BDL,branecos} but with an extra term contributed by the
Weyl tensor of the bulk. The conditions on the bulk spacetime geometry
required to recover the Einstein equations on the brane were studied in
ref.~\cite{DK}.
 
The simplest relevant model~\cite{RS2} consists of a single
hypermenbrane (i.e. brane of codimension one) with positive bare
tension located in a bulk spacetime that is asymptotically
anti-de-Sitter (adS) with reflection symmetry imposed along the extra
dimension {\it by hand}. Matter localized to the brane is assumed to be
a distributional source localised by a $\delta$-function.  It was shown in
ref.~\cite{RS2} that the gravitational fluctuations in this model, which
has an (effectively infinite) extra dimension, are those of a spin-two
graviton.  This model is not without its problems, for instance, the
apparent incompatibility with supersymmetry~\cite{susy} (see, however,
\cite{susy2} for supersymmetrisations of~\cite{RS1}) and instability to
unknown effects from the bulk~\cite{SMS,roy}, but it serves as an
interesting starting point for any investigation of the viability of
scenarios with extra dimensions.  Probing the assumptions of this model
is the basis of our discussion here.
 
Most of the works have focused on a 3-brane embedded in a 5 dimensional
spacetime with a reflection symmetry along the extra
dimension~\cite{RS2,BDL,branecos,SMS}. This was originally motivated by
the Ho\v{r}ava-Witten theory~\cite{HW} in which 11 dimensional M-theory
on an orbifold that is the product of a smooth manifold and $S^1/Z_2$
--- a circle with points identified by reflection symmetry about a
diameter.  This can be reduced to a 5 dimensional effective theory with
3-branes at the orbifold fixed points.  These 3-branes inherit gauge
groups from string theory which can produce the standard model gauge
group on one of the branes.  However, it should be emphasized that
$D$-branes do not have to be supported at an orbifold fixed
point~\cite{Polchinski}.

Although originally physically motivated, these two assumptions (of
dimensionality and reflection symmetry) owe their popularity to their
convenience. To derive the Einstein equations on the brane, one needs to
evaluate the extrinsic curvature of the brane in terms of the matter
content of the theory, which in the case of a 5 dimensional
$Z_2$-symmetric spacetime, is completely determined by the junction
conditions. The relevant Darmois-Israel junction conditions relate the
discontinuity of the extrinsic curvature to the matter content of the
brane (see, for example, \cite{SMS}), allowing the 4 dimensional
geometry of the brane to be calculated using the Gauss-Codacci
relations.  In more general cases where one has more transverse
dimensions the Gauss-Codacci equation relating the Ricci tensor of the
brane to the geometry of the bulk can be generalized (see
ref.~\cite{carter}) but there is no straightforward generalisation of
the Darmois-Israel junction conditions that apply to the hypersurface
supported case.  On the other hand, when the $Z_2$-symmetry hypothesis
is relaxed, the Darmois-Israel junction conditions \emph{alone} are not
sufficient to determine the extrinsic curvature, so one needs to
complement them with a dynamical equation for the brane
motion~\cite{carter,BC}.  Despite these technical problems, some models
with more than one extra dimensions have been studied~\cite{6D} as well
as some specific models in which the $Z_2$-symmetry was broken in an
{\it ad hoc} way~\cite{nonz2} by gluing two adS spacetimes with
different cosmological constants. Recently, a very natural mechanism for
breaking the $Z_2$-symmetry by means of a gauge form field has been
proposed~\cite{BC} and the Friedmann equations derived in this
particular case~\cite{CU}. We also note that some models with ``moving
branes''~\cite{moving} have been considered, which share some properties
with non-reflection symmetric brane model.
 
In this article, we relax the assumption of reflection symmetry and
investigate the gravitational effects experienced by observers on the
brane. To do this we choose the approach of constructing the effective
Einstein equation from the Gauss-Codacci equations, as done in
ref.~\cite{SMS} for the standard Randall-Sundrum model, and extended to
the more general case where there is a scalar field in the bulk in
refs~\cite{WM,MB}. This should be compared and contrasted with the
original~\cite{RS2} methods, which involve constructing an equation in
the linearized approximation for the so-called graviton wave function,
and higher energy modes due the effects of the extra dimension (see also
ref.~\cite{graviton}); such methods should be equally applicable in the
asymmetric case, but we do not consider them here.
  
As we noted above, the added complication in the case of an asymmetric
model is that more general junction conditions~\cite{BC} must be applied
to the extrinsic curvature at the brane, which take into account the
motion of the brane through the bulk spacetime. We first describe these
conditions for a co-dimension one brane-world in
section~\ref{sec-prelim}.  Using the Gauss equation, we shall show in
section~\ref{sec-series} that when the bare tension of the brane is
large compared to that the matter localized on the brane, the results of
ref.~\cite{SMS} can be generalized, with only modifications to the
effective cosmological and gravitational constant terms, and that in
principle higher order corrections can also be computed. This analysis
shows that the effective gravitational excitations in the weak field
limit are those of a massless spin-two graviton, as in the reflection
symmetric case, but with more complicated higher order contributions
from the bulk Weyl tensor. In section~\ref{sec-cod} we show that the
Codacci equation governs possible transfer of energy momentum 
between the brane and the bulk. In the usual cases, it implies that the
brane supported part of the energy momentum is conserved by itself.
 
Most work has focused on either the weak gravity as we do here, or the
cosmological case of a homogeneous and isotropic brane-world in which
one effectively probes gravity on large scales~\cite{branecos,BDL}. Even
when the $Z_2$ symmetry is broken, the dynamic equation for the
cosmological case can be solved exactly~\cite{CU}. In
section~\ref{sec-cos}, we show that the effective gravitational and
cosmological constants obtained in the perturbative analysis
carried out here agree with those obtained in the preceeding homogeneous
cosmological analysis~\cite{CU}.
 
In section~\ref{sec-form}, we consider the possibility that the absence
of reflection symmetry might result from the coupling of the brane to an
(antisymmetric) gauge form field that has bulk support. Outside the
distributional matter source on the brane, an $(n-1)$-form in an $n$
dimensional spacetime has no dynamics: its effect is just to modify the
cosmological constant in the bulk on either side, with a discontinuity
across the brane.

Whereas the earlier sections treat the general case in an $\dm$
dimensional bulk, finally, in section~\ref{sec-final} we consider the
specific case of a 3-brane coupled to a 4-form gauge field in a 5
dimensional bulk. We shall comment on the implications of our work for
linearized gravity and cosmological expansion.

\section{Hypermembrane dynamics}\label{sec-prelim} 
 
We shall consider the case of an $\dm-1$ dimensional brane in a 
$\dm$ dimensional spacetime (referred to as the bulk).  As standard matter 
(and thus observers) are confined to the brane in such models, the 
relevant geometry for comparison with observations will be that of the 
brane world-volume rather than the one of the bulk.  This will be given 
in terms of internal coordinates, $\sigma^a$, by embedding functions 
$x^\mu=\bar{x}^\mu(\sigma^a)$.  The induced brane world-volume metric, 
$\gamma_{ab}$, with respect to the internal coordinates can be expressed 
in terms of the bulk metric and embedding functions by 
\begin{equation} 
\gamma_{ab}=g_{\mu\nu}\frac{\partial\bar{x}^\mu}{\partial\sigma^a} 
\frac{\partial\bar{x}^\nu}{\partial\sigma^b} \,, 
\end{equation} 
where greek and latin indices run respectively from 0 to $\dm-1$ and from 
0 to $\dm-2$.  The contravariant inverse of $\gamma_{ab}$  and the associated 
internal Ricci tensor, $R^{ab}$, can be mapped into bulk tensors, given 
with respect to bulk coordinates, by 
\begin{equation} 
\ivm^{\mu\nu}=\gamma^{ab}\frac{\partial\bar{x}^\mu}{\partial\sigma^a} 
\frac{\partial\bar{x}^\nu}{\partial\sigma^b} \quad\mbox{and}\quad 
\brR^{\mu\nu}=R^{ab}\frac{\partial\bar{x}^\mu}{\partial\sigma^a} 
\frac{\partial\bar{x}^\nu}{\partial\sigma^b} \,, 
\end{equation} 
the first of which is known as the {\it first fundamental tensor} of the 
world-volume.  The corresponding mixed version ${\ivm^\mu}_\nu$ acts as 
a projection operator, projecting bulk tensors at a point on the brane 
into tensors tangential to the brane and defines a complementary 
orthogonal projection operator given by 
${\perp^\mu}_\nu\equiv{\delta^\mu}_\nu-{\ivm^\mu}_\nu$.  The above 
definitions are valid for a brane world-volume of arbitrary 
co-dimension but, in the case where the co-dimension is one, this 
orthogonal tensor can be expressed in terms of the unit normal, 
$\norm^\mu$ to the brane by ${\perp^\mu}_\nu=\norm^\mu\norm_\nu$. 
 
The internal Ricci tensor can be related to the Ricci tensor of the 
bulk, $\bkR_{\mu\nu}$, by the Gauss equation~\cite{carter} (see 
also ref.~\cite{schouten}) as  
\begin{equation} 
\label{ricci} 
\brR_{\mu\nu}=\frac{\dm-3}{\dm-2}\bkR_{\rho\s}{\ivm^\rho}_\mu{\ivm^\s}_\nu 
+\frac{1}{\dm-2}\bkR_{\rho\s}\ivm^{\rho\s}\ivm_{\mu\nu}-\frac{1}{\dm-1}\bkR\ivm_{\mu\nu}+ 
KK_{\mu\nu}-{K_\mu}^\rho K_{\nu\rho}+\bulk_{\mu\nu} \,, 
\end{equation} 
where $\bulk_{\mu\nu}$ is obtained by a surface projection from the 
Weyl tensor of the bulk spacetime, ${\cal C}_{\mu\rho\nu\s}$:
\begin{equation} 
\bulk_{\mu\nu}\equiv-{\cal C}_{\mu\rho\nu\s}\perp^{\rho\s} \,.
\end{equation} 
Since it is the trace free part of the bulk Riemann tensor, one can
show that 
\begin{equation} 
{\bulk^\mu}_\mu=0\,,\qquad 
\bulk_{\mu\nu}\perp^{\mu\nu}=0\,. 
\end{equation}

The extrinsic curvature $K_{\mu\nu}$ is defined as
\begin{equation} 
K_{\mu\nu}\equiv-{\eta^\s}_\nu{\eta^\alpha}_\mu\nabla_\alpha\norm_\s\,,
\end{equation} 
where $\nabla_\mu$ is the covariant derivative associated with 
$g_{\mu\nu}$.  Assuming that $\norm_\s$ is smoothly continued into the 
bulk along geodesics (i.e., that $\norm^\alpha\nabla_\alpha\norm_\mu=0$), 
it reduces to 
\begin{equation} 
K_{\mu\nu}=-\nabla_\mu\norm_\nu  
\end{equation} 
(using the conventions of ref.~\cite{MTW}). 
In a Gaussian normal coordinate system where the coordinate $\zeta$ 
measures the orthogonal distance from the brane (implying that 
$n_\mu=\nabla_\mu\zeta$), the background metric takes the standard form 
\begin{equation} 
\dd s^2_{_{[\dm]}}\equiv g_{\mu\nu}\dd x^\mu\dd x^\nu= 
\dd\zeta^2+\dd\bar s^2_{_{[\dm-1]}}\,,
\end{equation}  
with 
\begin{equation} 
\dd\bar s^2_{_{[\dm-1]}}\equiv\eta_{\mu\nu}\dd x^\mu\dd x^\nu= 
\gamma_{ab}\dd\s^a\dd\s^b \,,
\end{equation}  
and the extrinsic curvature is 
\begin{equation} 
2K_{\mu\nu}\dd x^\mu\dd x^\nu=-\frac{\dd}{\dd\zeta}\gamma_{ab}\dd\s^a\dd\s^b\,.
\end{equation} 

Although the bulk metric $g_{\mu\nu}$ is everywhere continuous, its 
derivative along $n^\mu$ may not be continuous across the brane. This
implies that $\bkR_{\mu\nu}$, $K_{\mu\nu}$ and
$\bulk_{\mu\nu}$ may also be discontinuous. We use 
square brackets to denote the jump of a particular quantity across 
the brane, defined with respect to the orientation of the normal $n_\mu$,
and angled brackets to denote its mean.  As an example, we decompose the 
extrinsic curvature tensor $K_{\mu\nu}$ as  
\begin{equation} 
\jump{K_{\mu\nu}}\equiv K^+_{\mu\nu}-K^-_{\mu\nu} \quad\mbox{and}\quad 
\avrg{K_{\mu\nu}}\equiv\frac{1}{2}\left(K^+_{\mu\nu}+K^-_{\mu\nu}\right) \,, 
\end{equation} 
where the $+$ superscript denotes the side of the brane into which the 
normal, $\norm_\mu$, points (the $\zeta>0$ region).  Since 
$\eta_{\mu\nu}$ must be smooth, $\brR_{\mu\nu}$ must be continuous 
across the brane so that the jump and average of the Gauss equation 
(\ref{ricci}) are given by
\begin{eqnarray} 
\label{ricciavrg} 
\avrg{\brR_{\mu\nu}}&=&\brR_{\mu\nu}=\avrg{\brF_{\mu\nu}}+\avrg{KK_{\mu\nu}} 
-\avrg{{K_\mu}^\rho K_{\nu\rho}} \,, \\ 
\label{riccijump} 
\jump{\brR_{\mu\nu}}&=&0=\jump{\brF_{\mu\nu}}+\jump{KK_{\mu\nu}} 
-\jump{{K_\mu}^\rho K_{\nu\rho}} \,, 
\end{eqnarray} 
where we have separated the contributions arising from the extrinsic
curvature and the ones from the bulk geometry, which are encapsulated
in $\brF_{\mu\nu}$ defined as
\begin{equation} 
\label{Fdef} 
\brF_{\mu\nu}\equiv\frac{\dm-3}{\dm-2}\bkR_{\rho\s}{\ivm^\rho}_\mu{\ivm^\s}_\nu 
+\frac{1}{\dm-2}\bkR_{\rho\s}\ivm^{\rho\s}\ivm_{\mu\nu}- 
\frac{1}{\dm-1}\bkR\ivm_{\mu\nu}+\bulk_{\mu\nu} \,. 
\end{equation} 
One can easily check that $\avrg{\;}$ and $\jump{\;}$ satisfy the 
two product relations 
\begin{equation} 
\label{product} 
\avrg{AB}=\avrg{A}\avrg{B}+\frac{1}{4}\jump{A}\jump{B} 
\quad \mbox{and} \quad 
\jump{AB}=\avrg{A}\jump{B}+\jump{A}\avrg{B} \,,
\end{equation} 
and, hence, to deduce $\brR_{\mu\nu}$ we need to compute
$\jump{K_{\mu\nu}}$ and $\avrg{K_{\mu\nu}}$ in terms of the stress
energy of the brane, $\overline{T}_{\mu\nu}$.
 
The extrinsic curvature of the brane is related to the stress energy
$\overline T^{\mu\nu}$ (with $\overline T^{\mu\nu}n_\mu=0$) supported on
the brane and to the force ($\bar f^\mu\equiv\bar f \norm^\mu$) exerted
on the brane by the bulk stress energy tensor by the generalisation of
Newton's second law~\cite{BC}
\begin{equation}\label{saileqn} 
\overline T^{\mu\nu}\avrg{K_{\mu\nu}}=\force \,, 
\end{equation} 
relating the contraction of the mean extrinsic curvature and
$\overline T^{\mu\nu}$ (reducing to mass times acceleration for a
point particle) to the force, $\force$, exerted by the bulk.  This
force is given in terms of the bulk energy-momentum tensor,
$T_{\text{ba}}^{\mu\nu}$, by 
\begin{equation}\label{saileqn2} 
\force=-\jump{T^{\mu\nu}_{\rm ba}}\norm_\mu\norm_\nu \,,
\end{equation}
which can be related to the bulk Ricci tensor $\bkR_{\mu\nu}$ by the $\dm$
dimensional Einstein equations.
We should note that the dynamical equation (\ref{saileqn}) can be
derived from (\ref{riccijump}), but that it has more general
applicability, for instance when gravity is not present, since it is
just a consequence of energy-momentum conservation. For our purposes,
we find that it is more economical to use (\ref{riccijump}).  In addition, the brane 
energy-momentum, $\overline T^{\mu\nu}$, determines the jump in the extrinsic 
curvature across the brane via the well-known junction 
conditions\footnote{These are almost universally referred to as the {\it 
Israel junction conditions}~\cite{israel}, but this extrinsic curvature 
method was used much earlier by the French mathematician George 
Darmois~\cite{darmois}.  Even earlier work on junctions in spacetimes 
was done by Lanczos~\cite{lanczos} and Sen~\cite{sen}. 
See~\cite{israel} for further references.} 
\begin{equation}\label{jumpcond} 
\jump{K\ivm_{\mu\nu}-K_{\mu\nu}}=\kap_{_{[\dm]}} \overline T_{\mu\nu} \,, 
\end{equation} 
where $\kap_{_{[\dm]}}$ is the gravitational constant of the 
$\dm$ dimensional spacetime related to the $\dm$ dimensional 
gravitational constant $G_{_{[\dm]}}$ by 
\begin{equation}\label{kappaG} 
\kap_{_{[\dm]}}\equiv(n-2)\Omega^{_{[\dm-2]}}G_{_{[\dm]}}\,,
\end{equation} 
and $\Omega^{_{[q]}}$ is the area of the $q$-sphere (for instance 
$\Omega^{_{[2]}}=4\pi$, $\Omega^{_{[3]}}=2\pi^2$\ldots). 
 
In brane-world models, it is common to interpret the stress energy of 
the brane as the sum of a bare tension and observed matter, the 
former being the dominant contribution (except, perhaps, in the very 
early universe) so that we shall decompose it as 
\begin{equation} 
\label{emsplit} 
\overline T_{\mu\nu}=-\tn\ivm_{\mu\nu}+{\overline{\cal T}}_{\mu\nu}= 
\tn(\emt_{\mu\nu}-\ivm_{\mu\nu}) \,, 
\end{equation} 
where $\tn(\ne 0)$ is the bare tension of the brane, and 
$\emt_{\mu\nu}\equiv{\overline{\cal T}}_{\mu\nu}/\tn$ is the dimensionless, 
energy-momentum localized to the brane.
In section \ref{sec-series} we will assume that
$\|\emt_{\mu\nu}\|\ll1$ to make a series expansion.
The jump condition (\ref{jumpcond}) inverts to give us
\begin{equation} 
\label{israel} 
\jump{K}=\kap_{_{[\dm]}}\tn\left(1+\frac{1-\emt}{\dm-2}\right)  
\quad\mbox{and}\quad 
\jump{K_{\mu\nu}}=\kap_{_{[\dm]}}\tn\left(\emt_{\mu\nu} 
+\frac{1-\emt}{\dm-2}\ivm_{\mu\nu}\right) \,, 
\end{equation}  
where $\emt={\emt^\mu}_\mu$ and the brane dynamical equation 
(\ref{saileqn}) becomes 
\begin{equation} 
\label{sail} 
\tn(\emt^{\mu\nu}-\ivm^{\mu\nu})\avrg{K_{\mu\nu}}=\force 
\qquad\hbox{with}\qquad 
2\kap_{_{[\dm]}}\force 
=\jump{2\bkR^{\mu\nu}\ivm_{\mu\nu}-\bkR} 
=\jump{\brF} \,.
\end{equation}
Here, we have made use of the bulk Einstein equations and the 
definition (\ref{Fdef}) of $\brF_{\mu\nu}$ to express the force,
$\force$, in terms of $\jump{\brF}$. 
Indeed, this equation is a scalar equation and alone cannot completely 
determine $\avrg{K_{\mu\nu}}$ therefore, as stressed above, we also have to 
use equation (\ref{riccijump}) which can be rewritten as 
\begin{equation}\label{1} 
\jump{\brF_{\mu\nu}}=\avrg{{K_\mu}^\rho}\jump{K_{\rho\nu}}+ 
\jump{{K_\mu}^\rho}\avrg{K_{\rho\nu}}-\avrg{K}\jump{K_{\mu\nu}} 
-\jump{K}\avrg{K_{\mu\nu}}\,. 
\end{equation} 
The general brane Einstein equation is then given by 
\begin{eqnarray}\label{2} 
\brR_{\mu\nu}&=&\frac{\dm-3}{\dm-2}\avrg{\bkR_{\rho\s}} 
             {\ivm^\rho}_\mu{\ivm^\s}_\nu 
             +\frac{1}{\dm-2}\avrg{\bkR_{\rho\s}}\ivm^{\rho\s}\ivm_{\mu\nu} 
             -\frac{1}{\dm-1}\avrg{\bkR}\ivm_{\mu\nu} 
             +\avrg{\bulk_{\mu\nu}} 
             +\frac{1}{4}\left(\jump{K}\jump{K_{\mu\nu}}- 
             \jump{{K_\mu}^\rho}\jump{K_{\nu\rho}} 
             \right)\nonumber\\ 
             &&+\avrg{K}\avrg{K_{\mu\nu}}-\avrg{{K_\mu}^\rho}\avrg{K_{\nu\rho}} 
             \,. 
\end{eqnarray} 
The equations we have stated allow us, in principle, to get expression 
for the $\dm-1$ dimensional Ricci scalar in terms of the matter in 
the spacetime, that is, both the matter supported throughout the bulk 
and that supported on the brane. In the most general case, equation 
(\ref{riccijump}) is not easy to invert.  In subsequent sections we 
solve this problem in two ways: firstly, in the case of arbitrary 
matter content, by a series expansion in $\emt_{\mu\nu}$ and,
secondly, in the case of cosmological models where the brane matter is
a perfect fluid and for which an exact solution can be found. 
 
Before trying to disentangle this set of equations, let us point out 
that imposing the reflection symmetry implies that 
$\jump{K_{\mu\nu}}=2K_{\mu\nu}$ and $\avrg{K_{\mu\nu}}=0$ which makes 
equation (\ref{sail}) and (\ref{1}) trivial and allows to directly 
express $\jump{K_{\mu\nu}}$ in terms of the brane stress-energy tensor 
$\overline T_{\mu\nu}$ by {\it only} making use of the Israel junction 
conditions (\ref{jumpcond}). The general brane equation (\ref{2}) then 
reduces to the one derived in ref.~\cite{SMS}. 
 
As a first interpretation of equation (\ref{2}), let us note that 
$\brR_{\mu\nu}$ has three origins: (i) an ``induced'' part coming from the 
geometry of the bulk, depending on the bulk Ricci and Weyl tensors,
(ii) a part coming from the jump in the extrinsic curvature across the
brane, and (iii) a part coming from the average of the extrinsic
curvature at the brane.  In the reflection symmetric models, only (i)
and (ii) contribute but, in the most general case, there is a
combination of all these effects.  In a companion paper~\cite{simulated},
we shall investigate the other extreme limiting case
were only (i) and (iii) have an effect.
 
\section{Series solution for general matter content} 
\label{sec-series} 
 
We are now in a position to obtain a series solution for $K_{\mu\nu}$ in 
terms of $\emt_{\mu\nu}$ using (\ref{riccijump}) by means of successive 
approximations, or equivalently expanding to the relevant order and 
separating by order.  The product relations (\ref{product}) and the 
Israel conditions (\ref{israel}) allow us to rewrite (\ref{riccijump}) 
as 
\begin{equation}\label{Rjump} 
\frac{1}{\kap_{_{[\dm]}}\tn}\jump{\brF_{\mu\nu}}=2{\emt_{(\mu}}^\rho\avrg{K_{\nu)\rho}}- 
\left(1-\frac{1-\emt}{\dm-2}\right)\avrg{K_{\mu\nu}} 
-\avrg{K}\left(\emt_{\mu\nu}+ 
\frac{1-\emt}{\dm-2}\ivm_{\mu\nu}\right)\,. 
\end{equation} 
So, if we expand $\avrg{K_{\mu\nu}}$ as 
\begin{equation} 
\label{series} 
\avrg{K_{\mu\nu}}=\avrg{K^{(0)}_{\mu\nu}}+\avrg{{K^{(1)}_{\mu\nu}}} 
+\order{\emt^2} \,, 
\end{equation} 
where $K^{(p)}_{\mu\nu}$ is of order ${\cal O}(\emt^p)$ in $\emt_{\mu\nu}$, 
and then substitute into (\ref{Rjump}) then the zeroth order term gives 
\begin{equation}\label{K0} 
\avrg{K^{(0)}_{\mu\nu}}=-\frac{1}{\kap_{_{[\dm]}}\tn(\dm-3)}\left( 
(\dm-2)\jump{\brF_{\mu\nu}}-\frac{1}{2}\jump{\brF}\ivm_{\mu\nu}\right)\,. 
\end{equation} 
Substituting this back into (\ref{Rjump}), the first-order 
term gives 
\begin{eqnarray}\label{K1} 
\avrg{{K^{(1)}_{\mu\nu}}}={1\over\kap_{_{[\dm]}}\tn(\dm-3)^2} 
\bigg(&&(\dm-2)\emt^{\rho\s}\jump{\brF_{\rho\s}}\ivm_{\mu\nu}+ 
(\dm-2)\emt\jump{\brF_{\mu\nu}}-2(\dm-2)^2{\emt_{(\mu}}^\rho 
\jump{\brF_{\nu)\rho}} \nonumber \\ 
&&{}+\frac{1}{2}(\dm-1)(\dm-2)\jump{\brF}\emt_{\mu\nu}- 
\frac{1}{2}(\dm-1)\emt\jump{\brF}\ivm_{\mu\nu}\bigg)\,. 
\end{eqnarray} 
This process could, of course, be iterated to calculate higher order 
terms of the series expansion (\ref{series}).  Motivated by the fact 
that $\bulk_{\mu\nu}$ is traceless, we decompose the bulk effects, 
$\brF_{\mu\nu}$, into the trace and a trace-free part, 
$\tfree_{\mu\nu}$ (${\tfree^\mu}_\mu=0$), so that 
\begin{equation} 
\label{Fsplit} 
\jump{\brF_{\mu\nu}}= 
       \frac{2\kap_{_{[\dm]}}\force}{\dm-1}\ivm_{\mu\nu} 
       +\kap_{_{[\dm]}}\jump{\tfree_{\mu\nu}} \,, 
\end{equation} 
from which, using expression (\ref{sail}) one can deduce that  
\begin{equation} 
\kap_{_{[\dm]}}\jump{\tfree_{\mu\nu}}=  
     \frac{\dm-3}{\dm-2}\jump{\bkR_{\rho\s}}\left( 
     {\ivm^\rho}_\mu{\ivm^\s}_\nu-\frac{1}{\dm-1} 
     \ivm^{\rho\s}\ivm_{\mu\nu}\right) 
     +\jump{\bulk_{\mu\nu}} \,. 
\end{equation} 
Equations (\ref{K0}) and (\ref{K1}) then simplify to 
\begin{eqnarray}\label{Kterms} 
\tn\avrg{K^{(0)}_{\mu\nu}}&=& -\frac{\force}{(\dm-1)}\ivm_{\mu\nu}- 
                          \frac{\dm-2}{\dm-3}\jump{\tfree_{\mu\nu}} \,, \\ 
\tn\avrg{{K^{(1)}_{\mu\nu}}}&=&\frac{\force}{(\dm-1)} 
                          \bigg((\dm-2)\emt_{\mu\nu}-\emt 
                          \ivm_{\mu\nu}\bigg)+\frac{\dm-2}{(\dm-3)^2} 
                          \bigg(\emt^{\rho\s}\jump{\tfree_{\rho\s}} 
                          \ivm_{\mu\nu}+\emt\jump{\tfree_{\mu\nu}}- 
                          2(\dm-2){\emt_{(\mu}}^\rho\jump{\tfree_{\nu) 
                         \rho}}\bigg) \,. 
\end{eqnarray} 
With this series expansion for $\avrg{K_{\mu\nu}}$, we can evaluate the 
Ricci tensor of the brane (\ref{2}) by substituting (\ref{israel}) and
(\ref{Kterms}) to eliminate the $K$ terms in favour of
$\emt_{\mu\nu}$, $\force$ and $\jump{\tfree_{\mu\nu}}$
\begin{eqnarray}\label{effRicci} 
\brR_{\mu\nu}&=&\left(\frac{1}{\dm-1}\avrg{\brF}+\frac{\kapp_{_{[\dm]}}\tn^2}{4(\dm-2)} 
+\frac{(\dm-2)\force^2}{(\dm-1)^2\tn^2}\right)\ivm_{\mu\nu}+ 
\bigg(\frac{\kapp_{_{[\dm]}}\tn^2}{4(\dm-2)}-{(\dm-2)\force^2\over 
(\dm-1)^2\tn^2} 
\bigg)\bigg((\dm-3)\emt_{\mu\nu}-\emt\ivm_{\mu\nu}\bigg)\nonumber\\ 
&& 
+\kap_{_{[\dm]}}\avrg{\tfree_{\mu\nu}}+{\force\over 
(\dm-1)\tn^2}\jump{\tfree_{\mu\nu}} 
-{(\dm-2)^2\over 
(\dm-3)^2\tn^2}\jump{{\tfree_{\mu}}^{\rho}}\jump{\tfree_{\rho\nu}}  
+{2(n-2)\force\over (\dm -1)(\dm 
-3)\tn^2}\bigg(2(\dm-2){\emt_{(\mu}}^{\rho} 
\jump{\tfree_{\nu)\rho}}-\emt\jump{\tfree_{\mu\nu}}\bigg)   
\nonumber\\ 
&& 
+{(\dm-2)^2\over 
(\dm-3)^3\tn^2}\bigg(2\emt\jump{{\tfree_{\mu}}^{\rho}} 
\jump{\tfree_{\rho\nu}}-4(\dm-2){\emt_{\rho}}^{\s} 
\jump{{\tfree_{(\mu}}^{\rho}}\jump{\tfree_{\nu)\s}}-4(\dm-2) 
{\emt_{(\mu}}^{\s}\jump{{\tfree_{\nu)}}^{\rho}}\jump{\tfree_{\rho\sigma}} 
-(\dm-5)\emt^{\rho\s}\jump{\tfree_{\rho\s}}\jump{\tfree_{\mu\nu}} 
\bigg) 
\nonumber\\ 
&& 
+\order{\emt^2}\,, 
\end{eqnarray} 
and, taking the trace, we compute the Ricci scalar to be 
\begin{eqnarray} 
\label{effR} 
\brR=&&\bigg(\avrg{\brF}+\frac{\kapp_{_{[\dm]}}\tn^2(\dm-1)}{4(\dm-2)}+\frac{(\dm-2)\force^2}{(\dm-1)\tn^2}\bigg) 
+\bigg(2{(\dm-2)\force^2\over(\dm-1)^2\tn^2}-\frac{\kapp_{_{[\dm]}}\tn^2}{2(\dm-2)}\bigg)\emt - {(\dm-2)^2\over 
(\dm-3)^2\tn^2}\jump{\tfree_{\rho\s}}\jump{\tfree^{\rho\s}} 
\nonumber\\ 
&& 
+{4(\dm-2)^2\force\over 
(\dm-1)(\dm-3)\tn^2}\emt_{\rho\s}\jump{\tfree^{\rho\s}} 
+{2(\dm-2)^2\over(\dm-1)^2\tn^2}\bigg(\emt\jump{\tfree_{\rho\s}}\jump{\tfree^{\rho\s}}- 
4(\dm-2){\emt_{\nu}}^{\rho}\jump{\tfree^{\nu\s}}\jump{\tfree_{\rho\s}}\bigg)+\order{\emt^2} 
\,. 
\end{eqnarray} 
Now, we can deduce the Einstein tensor
\begin{equation} 
\label{effEin} 
\brG_{\mu\nu}=-\Lambda_{_{[\dm-1]}}\ivm_{\mu\nu}+\kappa_{_{[\dm-1]}} 
{\overline{\cal T}}_{\mu\nu}+\order{\tfree,\emt^2}\,, 
\end{equation} 
where the cosmological constant on the brane is given by 
\begin{equation}\label{cosconst} 
\Lambda_{_{[\dm-1]}}={\dm-3\over 2}\bigg({\avrg{\brF}\over 
n-1}+{\kappa_{_{[\dm]}}^2\tn^2\over 
4(\dm -2)}+{(\dm-2)\force^2\over(\dm-1)^2\tn^2}\bigg)\,, 
\end{equation} 
and the gravitational constant, is identified to be 
\begin{equation}\label{gravconst} 
\kappa_{_{[\dm-1]}}=(\dm-3)\bigg( 
{\kappa_{_{[\dm]}}^2\tn\over 4(\dm -2)}- 
{(\dm-2)\force^2\over(\dm-1)^2\tn^3}\bigg)\,. 
\end{equation} 
It should be noted that these are exactly the same as found
ref.~\cite{CU} for the exact cosmological case with $\dm=5$, an 
issue to which we will return later. 
 
This enables us to compare these brane-world theories with more 
traditional gravity theories, such as four dimensional general 
relativity or Brans-Dicke theory. From the brane Einstein equation 
(\ref{effEin}) we can recover the results of ref.~\cite{SMS} in the
physically interesting case of $\dm=5$ by setting $\force$ to zero. As
with the reflection symmetric case, unequivocal 
interpretation of the result is problematic due to the contribution from 
bulk effects $\avrg{\tfree_{\mu\nu}}$, which is indeterminate from the 
point of view of observers on the brane. This is the case even in the 
reflection symmetric case~\cite{SMS,roy}; their interpretation being that 
these degrees of freedom correspond to the massive states of the higher 
dimensional graviton.  In the more general case studied here, there is 
an additional indeterminacy arising from $\jump{\tfree_{\mu\nu}}$ terms, 
but we should note that this is no more than one would expect since one 
will have massive modes from both sides. In this respect, the asymmetric 
case is at least no worse than that with reflection symmetry. 
 
\section{Codacci equation}\label{sec-cod} 
 
In addition to the Gauss equation the extrinsic 
curvature satisfies the Codacci equation, 
\begin{equation} 
\label{codacci} 
\brdel_\nu K-\brdel_\mu{K^\mu}_\nu=\bkR_{\rho\s}\norm^\s{\ivm^\rho}_\nu \,, 
\end{equation} 
where $\brdel_\mu$ denotes the covariant derivative associated to the 
induced brane metric $\ivm_{\mu\nu}$. 
Using the jump conditions (\ref{israel}), this equation 
gives an equation for the divergence of the brane energy-momentum 
$\emt_{\mu\nu}$  
\begin{equation} 
\label{emeqn} 
\kap_{_{[\dm]}}\tn\brdel_\mu{\emt^\mu}_\nu=-\jump{\bkR_{\rho\s}} 
\norm^\s{\ivm^\rho}_\nu \,. 
\end{equation} 
This relates the four dimensional divergence of $\emt_{\mu\nu}$ to the 
energy flow between the bulk and the brane, which is zero in the reflection
symmetric case, but not necessarily so in general. Since energy is 
observed to be conserved to a high degree in the late universe this will 
place strong restrictions on the relevant components of the bulk Ricci 
tensor. However we should note that violating brane energy conservation 
in the early universe may indeed be desirable in order to alleviate 
cosmological puzzles such as the horizon and flatness problems.  If the 
bulk spacetime is effectively dictated by two different bulk 
cosmological constants, as is the case when the asymmetry in the 
spacetime is generated by the form fields discussed in 
section~\ref{sec-form}, then the right hand side of (\ref{emeqn}) is zero 
and standard energy conservation on the brane is generic. 
 
There is also an equation coming from the average  of (\ref{codacci}), 
which restricts the variation of $\jump{\tfree_{\mu\nu}}$, in a 
similar way to the Bianchi identities constrain $\avrg{\tfree_{\mu\nu}}$. 
Using our series solution from Section~\ref{sec-series}, the zeroth 
order terms of this give us 
\begin{equation} 
\frac{1}{\tn}\left(\frac{1}{2(\dm-1)}\brdel_\nu\force+ 
\frac{\dm-2}{\dm-3}\brdel_\mu\jump{{\tfree^\mu}_\nu}\right) 
=\avrg{\bkR_{\rho\s}}\norm^\s{\ivm^\rho}_\nu \,. 
\end{equation} 
Clearly, the first order corrections will be constrained by similar relations.

\section{Exact solution for cosmological models}\label{sec-cos} 
 
Of great interest are cosmological models which are  
spatially isotropic and homogeneous.  This means that the brane matter 
content will have the specific isotropic and homogeneous form 
\begin{equation} 
\label{fluid} 
{\overline{\cal T}}_{\mu\nu}= 
           {\cal U}u_{\mu}u_{\nu}-{\cal T}(u_\mu u_\nu+\ivm_{\mu\nu}) 
          =\tn\left[\den u_\mu u_\nu+\prs(\ivm_{\mu\nu}+u_\mu u_\nu) 
          \right]\,, 
\end{equation} 
where $\den$ and $\prs$ are, respectively, the dimensionless ratios of 
the energy density and pressure to the bare brane tension, $\tn$, and $u_\mu$ 
is the unit timelike vector tangent to the brane 
(that is, $u^\mu u_\mu=-1$ and $\perp_{\mu\nu}u^\nu=0$)
which is normal to the $(n-2)$-surfaces 
of isotropy and homogeneity (that is, to the spatial sections of our 
universe). In this case, equation (\ref{Rjump}) becomes 
\begin{equation} 
\label{cosRjump} 
\frac{1}{\kap_{_{[\dm]}}\tn}\jump{\brF_{\mu\nu}}=\left(\frac{\den+1}{\dm-2}+\prs-1\right) 
\avrg{K_{\mu\nu}}+2(\den+\prs)\avrg{{K_{(\mu}}^\rho}u_{\nu)}u_\rho- 
\left(\frac{\den+1}{\dm-2}\ivm_{\mu\nu}+(\den+\prs)u_\mu u_\nu\right) 
\avrg{K}\,. 
\end{equation} 
This can be inverted exactly (see appendix~\ref{app-invert} for details) 
to give 
\begin{eqnarray} 
\label{Kcos1} 
\kap_{_{[\dm]}}\tn\left(\frac{\den+1}{\dm-2}+\prs-1\right)\avrg{K_{\mu\nu}}&=& 
\jump{\brF_{\mu\nu}}+2\frac{(\den+\prs)(\dm-2)}{(\den+1)(\dm-3)} 
\jump{{\brF_{(\mu}}^\rho}u_{\nu)}u_\rho+\frac{(\dm-2)(\den+\prs)^2}{(\dm-3)(\den+1)^2} 
\jump{\brF_{\rho\s}}u^\rho u^\s u_\mu u_\nu \nonumber \\ 
&&{}-\frac{(\den+\prs)(\den-(\dm-2)\prs+\dm-1)}{2(\den+1)^2(\dm-3)} 
\jump{\brF}u_\mu u_\nu \nonumber \\ 
&&{}-\left(\frac{\den+\prs}{(\dm-3)(\den+1)}\jump{\brF_{\rho\s}}u^\rho u^\s+ 
\frac{(2\dm-5)\den+(\dm-2)\prs+\dm-3}{2(\den+1)(\dm-2)(\dm-3)}\jump{\brF}\right) 
\ivm_{\mu\nu} \,. 
\end{eqnarray} 
Making the split (\ref{Fsplit}) of $\brF_{\mu\nu}$ into its trace, 
$\force$, and its trace-free part, $\tfree_{\mu\nu}$, we get 
\begin{eqnarray} 
\label{Kcos} 
\tn\avrg{K_{\mu\nu}}&=&\frac{\force}{(\dm-1)(\den+1)^2} 
\left((\den+\prs)(\dm-2)u_\mu u_\nu-(\den+1)\ivm_{\mu\nu}\right) 
+\left(\jump{\tfree_{\mu\nu}}+2\frac{(\den+\prs)(\dm-2)}{(\den+1)(\dm-3)} 
\jump{{\tfree_{(\mu}}^\rho}u_{\nu)}u_\rho \right. \nonumber \\ 
&&\left.{}+\frac{(\dm-2)(\den+\prs)^2}{(\dm-3)(\den+1)^2} 
\jump{\tfree_{\rho\s}}u^\rho u^\s u_\mu u_\nu 
-\frac{(\den+\prs)}{(\dm-3)(\den+1)}\jump{\tfree_{\rho\s}} 
u^\rho u^\s\ivm_{\mu\nu}\right)\frac{\dm-2}{\den+(\dm-2)\prs-\dm+3} \,, 
\end{eqnarray} 
and the trace of (\ref{Kcos}) has the relatively simple form 
\begin{equation} 
\tn\avrg{K}=\frac{(\dm-2)(\den+\prs)-(\dm-1)(\den+1)}{(\dm-1)(\den+1)^2}\force- 
\frac{(\den+\prs)(\dm-2)}{(\dm-3)(\den+1)^2} 
\jump{\tfree_{\rho\s}}u^\rho u^\s \,.
\end{equation} 
 
Equation (\ref{Kcos}) can be substituted into (\ref{2}), along with the 
junction conditions, to give an exact expression for the 
$\dm-1$ dimensional Einstein tensor.  As we have spatial isotropy and 
homogeneity, there is only one free Einstein equation, which it is 
convenient to take as the equation for the Ricci scalar, given by the 
trace of (\ref{2}) as 
\begin{equation}
\label{cosmoRicci}
\brR=\avrg{\brF}+\frac{1}{4}\left(\jump{K}^2-\jump{K^{\mu\nu}}\jump{K_{\mu\nu}}\right) 
+\avrg{K}^2-\avrg{K^{\mu\nu}}\avrg{K_{\mu\nu}} \,. 
\end{equation} 
So the part of the Ricci scalar which does not depend on 
$\tfree_{\mu\nu}$ is 
\begin{eqnarray} 
\brR&=&\avrg{\brF}+\frac{\kapp_{_{[\dm]}}\tn^2}{4(\dm-2)}\bigg(\dm-1+ 
2\big(\den-(\dm-2)\prs\big)-\big(2(\dm-2)\prs+ 
(\dm-3)\den\big)\den\bigg)\nonumber\\ 
&&+\frac{(n-2)\force^2}{\tn^2(\dm-1)(\den+1)^3}\bigg((n-1)(\den+1)+2(\dm-2) 
(\den+\prs) \bigg) \,. 
\end{eqnarray} 
It is simple to check for consistency with (\ref{effR}) by expanding the 
denominator of the last term keeping only the terms linear in $\den$ and 
$\prs$. One finds that the relevant coefficients agree completely with 
the definitions of $\Lambda_{_{[n-1]}}$ and $\kappa_{_{[n-1]}}$ given by 
(\ref{cosconst}) and (\ref{gravconst}). Hence, we have confirmed the 
results of section~\ref{sec-series} are correct to linearized order for 
an isotropic perfect fluid.

\section{Specific case of a bulk form field}\label{sec-form} 
 
We now discuss a simple possible realization of our asymmetric 
brane-world scenario in terms of an $(\dm-1)$-form field 
${A^{\{\dm-1\}}}_{\nu_1...\nu_{\dm-1}}$ which might naturally arise in 
some M-theory compactification which gives rise to the model. Such 
field couple to the brane world-volume via a generalized Wess-Zumino 
interaction and have field strength 
\begin{equation} 
{F^{\{\dm\}}}_{\nu_0\nu_1...\nu_{\dm-1}}=\dm\nabla_{[\nu_0} 
{A^{\{\dm-1\}}}_{\nu_1...\nu_{\dm-1}]}\,. 
\end{equation} 
As discussed in ref.~\cite{BC}, this field strength can be 
written in terms of a single pseudo-scalar $F^{\{\dm\}}$ and the totally 
antisymmetric tensor $\epsilon_{\nu_0...\nu_{\dm-1}}$.  Moreover, the dynamics 
of this field are such that it will be constant except for jump at the 
brane which is proportional to their coupling, $e^{\{\dm\}}$, when the 
matter on the brane is $\delta$-function localized. Hence, effectively,
one has a scenario with a discontinuous bulk cosmological constant such 
that 
\begin{equation} 
\force=e^{\{\dm-1\}}\avrg{F^{\{\dm\}}}=\kappa_{_{[\dm]}}^{-1} 
\jump{\Lambda_{_{[\dm]}}}\,. 
\end{equation} 
 
If no other matter is present in the bulk, its Ricci tensor will then 
satisfy the following minimal coupling isotropy condition, 
\begin{equation} 
\bkR_{\mu\nu}=\frac{1}{\dm}\bkR g_{\mu\nu} \,, 
\end{equation} 
where $\bkR$ will necessarily be uniform on each side 
of the brane.  This gives 
\begin{equation} 
\brF_{\mu\nu}=\frac{\dm-2}{\dm(\dm-1)}\bkR\ivm_{\mu\nu}+\bulk_{\mu\nu} \,. 
\end{equation} 
Thus equation (\ref{Fsplit}) implies that 
\begin{equation} 
\omega_{\mu\nu}=\frac{\bulk_{\mu\nu}}{\kap_{_{[\dm]}}}\qquad\hbox{and}\qquad 
\brF=\frac{\dm-2}{\dm}\bkR\,. 
\end{equation} 
Hence the right hand side of the four dimensional energy 
conservation equation (\ref{emeqn}) is zero so, if we consider the case 
where the bare tension, $\tn$, is constant, then we still have 
conservation of the energy-momentum tensor $\emt_{\mu\nu}$. 
 
We now consider cosmological solutions where the bulk has 
$(n-2)$ dimensional spatial surfaces of isotropy and homogeneity.  With 
this minimally coupled form, it is possible to apply a generalized 
Birkhoff theorem to such a spacetime, as was demonstrated in
ref.~\cite{BCG} and applied to this situation in ref.~\cite{CU}.
According to this theorem, the metric in the bulk will
not just be homogeneous and isotropic but necessarily static, so that in 
suitable coordinates it will be expressible in the the form 
\begin{equation} 
\label{staticansatz} 
\dd s^2_{_{[\dm]}}=r^2\dd\ell^2+\frac{\dd r^2}{\cal V}-{\cal V}\dd t^2 \,, 
\end{equation} 
where $\dd\ell^2$ is the (positive-definite) space metric of an 
$\dm-2$ dimensional sphere, plane or anti-sphere with constant 
curvature, $k$ say --- which will be respectively positive, zero, or 
negative --- and ${\cal V}$ is given on either side by an expression of 
the generalized Schwarzschild-de Sitter form 
\begin{equation} 
\label{Schw} 
{\cal V}=k-\frac{\bkR^{^\pm}}{\dm(\dm-1)}r^2 
-\frac{2G_{_{[\dm]}}{\cal M}^{^\pm}}{r^{\dm-3}}\,. 
\end{equation} 
where ${\cal M}^{_+}$ and ${\cal M}^{_-}$ are constants, which 
will be interpretable, in the case of positive curvature with the 
standard normalization $k=1$, as representing the total asymptotic 
mass on either side. 
 
As discussed in ref.~\cite{CU}, in this homogeneous and 
isotropic case, the line element can be re-written in terms of Gaussian 
normal coordinates as 
\begin{equation} 
\label{GNansatz} 
\dd s^2_{_{[\dm]}}=r^2\dd\ell^2+\dd\zeta^2-\nu^2 \dd\tau^2 \,, 
\end{equation} 
where the brane is located at $\zeta=0$ and the new time coordinate 
$\tau$ is constant on geodesics orthogonal to the brane. 
The function $\nu$ depends only on $\zeta$ and $\tau$, and can be
normalized so that $\nu\rightarrow1$ when $\zeta \rightarrow 0$ --- that
is, on the brane.  This means that $\tau$ will simply be a measure of
proper time within the brane.  The corresponding limit
$r(\tau,\zeta)\rightarrow a(\tau)$ defines a function,
interpretable as a cosmological scale factor, which satisfies
\begin{equation}\label{aeqn}  
\left(\frac{\dd a}{\dd\tau}\right)^2= 
\left(\frac{\partial\alpha}{\partial\zeta}\right)^2-{\cal V}\,.
\end{equation} 
In addition, $\nu$ must satisfy 
\begin{equation} 
\nu \frac{\dd a}{\dd\tau}=\frac{\partial r}{\partial\tau}\,. 
\end{equation} 
It follows that the first fundamental form, $\ivm_{\mu\nu}$, is 
given in this case by 
\begin{equation} 
\label{braneansatz} 
\dd\bar{s}^2_{_{[\dm-1]}}=a^2\dd\ell^2-\dd\tau^2\,, 
\end{equation} 
and the second fundamental form, $K_{\mu\nu}$, is given by 
\begin{equation} 
\label{sffansatz} 
K_{\mu\nu}\,\dd x^\mu \dd x^\nu =-a\frac{\partial r}{\partial\zeta} 
\dd\ell^2+\frac{\partial\nu}{\partial\zeta}\dd\tau^2\,. 
\end{equation} 
 
Following the method described in ref.~\cite{MTW}, it is 
possible, in the case of a spacetime with the line element given by 
(\ref{staticansatz}) and (\ref{Schw}), to derive an expression for the 
contribution of the bulk projected Weyl tensor, $\bulk_{\mu\nu}$, which 
can be expressed as (see Appendix~\ref{app-weyl} for details) 
\begin{equation} 
\label{weylfield} 
\bulk_{\mu\nu}\,\dd x^\mu\dd x^\nu={\cal W}^{^\pm} 
\left(a^2\dd\ell^2+(\dm-2)\dd\tau^2\right) \,, 
\end{equation} 
in which the Weyl scalars, ${\cal W}^{\pm}$ --- which, of course, are not the 
trace of the Weyl tensor since this that would give zero --- are related 
to the asymptotic mass on each side of the brane by 
\begin{equation} 
\label{weylscalar} 
{\cal W}^{^\pm}=(\dm-3)\frac{G_{_{[\dm]}}{\cal M}^{^\pm}}{a^{\dm-1}} \,. 
\end{equation} 
Expression (\ref{weylscalar}) leads to the usual interpretation of
the Weyl contribution as a pseudo-radiation term when $n=5$.
In terms of the preferred timelike unit vector, $u_\mu=-\nabla_\mu\tau$, 
used in the definition of (\ref{fluid}), which specifies the 
cosmological rest frame, $\bulk_{\mu\nu}$ can be written as
\begin{equation} 
\bulk_{\mu\nu}={\cal W}^{^\pm}\bigg(\ivm_{\mu\nu}+(\dm-1)u_\mu u_\nu\bigg)\,, 
\end{equation} 
and, therefore, substituting this into (\ref{Kcos}) gives us the
expression for $\avrg{K_{\mu\nu}}$,
\begin{eqnarray} 
\tn\avrg{K_{\mu\nu}}&=&\frac{\force}{(\dm-1)(\den+1)^2} 
\bigg((\den+\prs)(\dm-2)u_\mu u_\nu-(\den+1)\ivm_{\mu\nu}\bigg)\nonumber\\ 
&&{}-\frac{\jump{{\cal M}}}{\Omega^{[n-2]}a^{\dm-1}(\den+1)^2} 
\bigg(\big(\den+(\dm-2)\prs+\dm-1\big)u_\mu 
u_\nu+(\den+1)\ivm_{\mu\nu}\bigg)\,,  
\end{eqnarray} 
which depends only on $\den$, $\prs$ and on the two constant parameters 
$\force$ and $\jump{{\cal M}}$.  The trace of these equation has the 
simple form 
\begin{equation} 
\tn\avrg{K}=\frac{1-\dm-(2\dm-3)\den-(\dm-2)\prs}{(\dm-1)(\den+1)^2}\force 
-\frac{(\dm-2)(\den+\prs)}{\Omega^{[\dm-2]}(\den+1)^2}\jump{{\cal M}} \,. 
\end{equation} 
These equations can then be substituted into (\ref{1}) and (\ref{2}) to
derive the results found in ref.~\cite{CU}. We have shown that, in the
perfectly homogeneous case, $\avrg{\tfree_{\mu\nu}}\propto\avrg{\cal
M}$ and $\jump{\tfree_{\mu\nu}}\propto\jump{\cal M}$, where ${\cal M}$
is the mass on any Schwarzschild black hole which resides in the bulk.
In the original Randall-Sundrum scenario~\cite{RS2} the bulk spacetime
is pure adS, so ${\cal M}$ would be zero.
 
\section{Summary: 3-brane in a five dimensional spacetime} 
\label{sec-final} 
 
In this article, we have obtained the effective Einstein equations for a 
$\dm-1$ dimensional hypermembrane embedded in a $\dm$ dimensional spacetime 
without imposing any reflection symmetry along the extra dimension, 
hence generalizing the previous work of ref.~\cite{SMS}. We have found
[see, for example, equation (\ref{2})] that the 
$\dm-1$ dimensional Ricci tensor has two origins, one coming from the bulk 
geometry and a second due to the breakdown of the  reflection symmetry.
In general, the closed set of equations is difficult to invert but we 
have shown that it is possible firstly, by means of a series expansion in the 
weak field limit in which the ordinary matter was a small contribution 
with respect to the brane tension and, secondly, exactly if the matter in the 
brane is a perfect fluid.

Now, to summarize our results let us consider the specific physically 
interesting case where $n=5$. We have shown is that in such a model the 
curvature tensor on the brane (\ref{effEin}) can be written in the 
simple form 
\begin{equation} 
\brG_{\mu\nu}=-\Lambda_{_{[4]}}\ivm_{\mu\nu}+8\pi G_{_{[4]}} 
{\overline{\cal T}}_{\mu\nu} \,,
\end{equation} 
if one ignores $\order{{\cal T}^2}$ terms and the effects of the bulk 
modes, which we have argued will be related to higher dimensional massive 
modes in an exactly similar way to the case when the system is 
reflection symmetric. The relevant cosmological (\ref{cosconst}) and 
gravitational (\ref{gravconst}) constants are given, whatever the matter 
on the brane, by 
\begin{equation}\label{l4} 
\Lambda_{_{[4]}}={\avrg{\brF}\over 4}+{\kap^2_{_{[5]}}\tn^2\over 
12}+{3\force^2\over 16\tn^2}\,, 
\end{equation} 
and  
\begin{equation} \label{g4}
\kap_{_{[4]}}=8\pi G_{_{[4]}}={\kap_{_{[5]}}^2\tn\over 6} 
-{3\force^2\over8\tn^3}\,, 
\end{equation} 
which are exactly the same as derived in the cosmological 
case~\cite{CU}. Moreover, we have shown that the precise form of the 
cosmological expansion rate is the same once it is linearized to the 
relevant order. Finally, we have pointed out that, in such a scenario, an 
asymmetric brane-world is natural within a model where 4-form field is 
coupled to the brane via a generalized Wess-Zumino action. This puts on 
a more solid footing the earlier, more {\it ad hoc} work on this 
reflection symmetry breaking scenarios~\cite{nonz2}. 
 
Introducing the two mass scales $m_{_\Lambda}^{^\pm}$, the inverse of 
which are length scales characterizing the bulk geometry on both sides, 
\begin{equation} 
m_{_\Lambda}^{^\pm}\equiv\pi G_{_{[5]}}\overline{\cal T}_{_\infty} 
\pm\frac{\force}{4\overline{\cal T}_{_\infty}}, 
\end{equation} 
we can rewrite (\ref{l4}-\ref{g4}) as 
\begin{equation} 
8\pi G_{_{[4]}}=\frac{6}{\overline{\cal T}_{_\infty}}m_{_\Lambda}^{^+} 
m_{_\Lambda}^{^-}\qquad\hbox{and}\qquad 
2\Lambda_{_{[4]}}=\frac{\avrg{{\cal F}}}{2}+3\avrg{m_{_\Lambda}^2}. 
\end{equation} 
In the particular case of a maximally symmetric background, which is the 
case when a 4-form field is coupled to the brane, ${\cal 
R}_{\mu\nu}={\cal R}g_{\mu\nu}/\dm$ and we obtain that $\avrg{{\cal F}}= 
(\dm-2)\avrg{{\cal R}}/\dm$ so that for a five dimensional anti-de-Sitter
spacetime $\avrg{{\cal F}}=2\avrg{\Lambda}$. The two mass scales 
$m_{_\Lambda}^{^\pm}$ need not only to be positive but large enough to 
lead to a viable cosmology~\cite{CU}. 
 
\vskip0.5cm  
The original result of Randall and Sundrum that five 
dimensional gravity can yields standard Einstein gravity on a brane with 
positive bare tension was a surprising one. It was mainly based on the 
fact that the extrinsic curvature was large enough to simulate the 
effect of a small compactification radius.  Here, we have shown it 
generalizes to a brane which is not reflection symmetric provide the 
curvature radius on {\it each side} is small enough or, equivalently,
that the mass scales, $m_{_\Lambda}^{^\pm}$, are both large enough ---
who knows where it will end!

\section*{Acknowledgements} 
 
\noindent RAB is funded by PPARC and AM is funded by EPSRC.
JPU would like to thank D. Steer and M. Ruiz-Altaba for discussions on
moving branes while he was in Sils Maria, as well as E. Dudas and J. Mourad.

\appendix 
\section{Calculating $\avrg{K_{\mu\nu}}$ for a perfect fluid} 
\label{app-invert} 
 
To invert equation (\ref{cosRjump}), we need to compute $\avrg{K}$ and 
$\avrg{K_{\mu\nu}}u^\mu$. The trace of (\ref{cosRjump}) and its 
contraction with $u^\mu u^\nu$ respectively give the system 
\begin{eqnarray} 
&&\frac{1}{\kap_{_{[\dm]}}\tn}\jump{\brF}=2(\prs-1)\avrg{K}+2(\den+\prs) 
\avrg{K_{\mu\nu}}u^\mu u^\nu \,, \\ 
&&\frac{\dm-2}{\kap_{_{[\dm]}}\tn}\jump{\brF_{\mu\nu}}u^\mu u^\nu= 
\left(1+\den-(\den+\prs)(\dm-2)\right)\avrg{K}+ 
\left(1+\den-(1+2\den+\prs)(\dm-2)\right)\avrg{K_{\mu\nu}}u^\mu u^\nu \,. 
\end{eqnarray} 
This system of equations can be solved to give expressions for 
$\avrg{K}$ and $\avrg{K_{\mu\nu}}u^\mu u^\nu$: 
\begin{eqnarray} 
\avrg{K}&=&-\frac{(\dm-2)(\den+\prs)}{\kap_{_{[\dm]}}\tn(\den+1)^2(\dm-3)} 
\jump{\brF_{\mu\nu}}u^\mu u^\nu-\frac{(2\dm-5)\den+(\dm-2)\prs+\dm-3} 
{2\kap_{_{[\dm]}}\tn(\den+1)^2(\dm-3)}\jump{\brF} \,, \\ 
\avrg{K_{\mu\nu}}u^\mu u^\nu&=&-\frac{(\dm-2)(1-\prs)}{\kap_{_{[\dm]}} 
\tn(\den+1)^2(\dm-3)} 
\jump{\brF_{\mu\nu}}u^\mu u^\nu+\frac{(\dm-3)\den+(\dm-2)\prs-1} 
{2\kap_{_{[\dm]}}\tn(\den+1)^2(\dm-3)}\jump{\brF} \,. 
\end{eqnarray} 
Substituting these two expressions back into the contraction of 
(\ref{cosRjump}) with $u^\nu$ gives us an expression for 
$\avrg{K_{\mu\nu}}u^\nu$ 
\begin{equation} 
-\avrg{K_{\mu\nu}}u^\nu=\frac{\dm-2}{\kap_{_{[\dm]}}\tn(\dm-3)(\den+1)} 
\jump{\brF_{\mu\nu}}u^\nu+\frac{(\dm-2)(\den+\prs)}{\kap_{_{[\dm]}}\tn(\dm-3) 
(\den+1)^2}\jump{\brF_{\rho\s}} 
u^\rho u^\s u_\mu+\frac{(\dm-3)\den+(\dm-2)\prs-1}{2\kap_{_{[\dm]}}\tn(\dm-3)(\den+1)^2}\jump{\brF}u_\mu \,. 
\end{equation} 
Substituting all of these back into (\ref{cosRjump}) gives (\ref{Kcos1}). 
 
\section{Weyl curvature for hyper-spherical spacetimes} 
\label{app-weyl} 
 
The goal of this appendix is to give the element of derivation of 
equation (\ref{weylscalar}) and to compute $\bulk_{\mu\nu}$ as given in 
(\ref{weylfield}).  For that purpose, we consider a $\dm$ dimensional 
Schwarzschild-de-Sitter bulk spacetime with line element
\begin{equation} 
\dd s^2_{_{[\dm]}}=-{\cal V}(r)\dd t^2+\frac{\dd r^2}{{\cal V}(r)}+r^2\left( 
\frac{\dd\rho^2}{1-k\rho^2}+\rho^2\dd\ell_{[\dm-3]}^2\right)\,, 
\end{equation} 
where $\dd\ell_{[\dm]}^2$ is the measure on the unit $\dm$-sphere. 
The problem is that the line element for the unit $(\dm-3)$-sphere is 
not easy to write in a compact form.  However, because of the symmetry 
of the spacetime, we need only to calculate the curvature at one point 
on the $(n-2)$-surfaces of spatial isometry.  Thus, we perform a 
coordinate transformation to make these $(n-2)$-surfaces manifestly 
conformally flat and then use Cartesian coordinates, which makes it easy 
to generalize to higher dimensions.  When $k=0$ this is trivial so we 
focus on the two cases $k=\pm1$, where we define a new 
coordinate $\chi=\chi(\rho)$ such that the line element of these 
$(n-2)$-surfaces is 
\begin{equation} 
\dd s_{_{[\dm-2]}}^2=r^2\left(\frac{\dd\rho^2}{1-k\rho^2}+\rho^2\dd 
\ell_{[\dm-3]}^2\right) 
=\frac{r^2}{\big(\chi'(\rho)\big)^2(1-k\rho^2)}\left(\dd\chi^2+ 
\rho^2(1-k\rho^2)\big(\chi'(\rho)\big)^2\dd\ell_{[\dm-3]}^2\right)\,, 
\end{equation} 
with prime denoting a derivative with respect to $\rho$.  Thus, we
need to choose $\chi(\rho)$ so that
\begin{equation}\label{ll} 
\rho^2(1-k\rho^2)\big(\chi'(\rho)\big)^2=\chi(\rho)^2 \,. 
\end{equation} 
For $k=\pm1$, equation (\ref{ll}) can be integrated to give 
\begin{equation}\label{coordxfm} 
\chi(\rho)=k\frac{1-\sqrt{1-k\rho^2}}{\rho}\,, 
\end{equation} 
It follows that the line element of the $(n-2)$-surfaces of spatial isotropy is 
\begin{equation}\label{openline} 
\dd s_{_{[n-2]}}^2=\frac{2r^2}{1+k\chi^2} 
\left(\dd\chi^2+\chi^2\dd\ell_{[\dm-3]}^2\right)\,, 
\end{equation} 
which is manifestly conformally flat.  Hence, writing (\ref{openline}) as the appropriate 
conformal factor times the usual line element for ${\Bbb R}^{n-2}$, we 
conclude that the full $n$ dimensional line element can be written 
\begin{equation} 
\label{SadSconf} 
\dd s^2_{_{[\dm]}}=-{\cal V}(r)\dd t^2+\frac{\dd r^2}{{\cal V}(r)}+ 
\frac{2r^2}{\Omega^2}\big((\dd x^1)^2+\cdots+(\dd x^{n-2})^2\big)\,, 
\end{equation} 
where $2\Omega^2\equiv1+k\big((x^1)^2+\cdots+(x^{n-2})^2\big)$, which holds 
also for $k=0$. 
 
We are now in a position to calculate the curvature tensors, which we do 
using the Cartan 2-form method (as outlined in ref.~\cite{MTW}).  We write 
the metric (\ref{SadSconf}) in terms of a basis of orthonormal 
1-forms as 
\begin{equation} 
\dd s^2=-\big(\bom^t\big)^2+\big(\bom^r\big)^2+ 
\big(\bom^1\big)^2+\cdots+\big(\bom^{(\dm-2)}\big)^2\,, 
\end{equation} 
where $\lbrace\bom^t,\bom^r,\bom^i\rbrace$ is 
explicitly given by 
\begin{equation} 
\bom^t\equiv{\cal V}^{1/2}\dd t \,,\quad 
\bom^r\equiv\frac{\dd r}{{\cal V}^{1/2}} \,, \quad 
\bom^i\equiv\frac{r}{\Omega}\dd x^i \,. 
\end{equation} 
Note that boldface letters denote 1-forms and the indices \emph{do 
not} denote components of these forms but labels. However, we will 
invoke the usual summation convention over repeated indices.  The 
exterior derivatives of these 1-forms are 
\begin{eqnarray} 
\dd\bom^r&=&0\\ 
\dd\bom^t&=&\frac{{\cal V}'}{2{\cal V}^{1/2}}\dd r\wedge \dd t 
=\frac{{\cal V}'}{2{\cal V}^{1/2}}\bom^r\wedge\bom^t \,, \\ 
\dd\bom^i&=&\frac{1}{\Omega}\dd r\wedge \dd x^i-\frac{r}{\Omega^2} 
\frac{\dd\Omega}{\dd x^j}\dd x^j\wedge \dd x^i=\frac{{\cal V}^{1/2}}{r} 
\bom^r\wedge\bom^i+\frac{1}{r}\frac{\dd\Omega}{\dd x^j} 
\bom^i\wedge\bom^j \,. 
\end{eqnarray} 
Using the notation of ref.~\cite{MTW}, we need to find the connection 
1-forms ${\bom^\mu}_\nu$ which are antisymmetric 
(that is, $\bom_{\mu\nu}=-\bom_{\nu\mu}$) and satisfy the first Cartan 
structure equation, $\dd\bom^\mu=-{\bom^\mu}_\nu\wedge \bom^\nu$, for a 
torsion-free case, which can easily be seen to be satisfied by 
\begin{eqnarray} 
{\bom^t}_r&=&{\bom^r}_t=\frac{{\cal V}'}{2{\cal V}^{1/2}}\bom^t 
=\frac{{\cal V}'}{2}\dd t \,, \\ 
{\bom^t}_i&=&{\bom^i}_t=0 \,, \\ 
{\bom^i}_r&=&-{\bom^r}_i=\frac{{\cal V}^{1/2}}{r}\bom^i 
=\frac{{\cal V}^{1/2}}{\Omega}\dd x^i \,, \\ 
{\bom^i}_j&=&\frac{1}{r}\left(\frac{\dd\Omega}{\dd x^i}\bom^j- 
\frac{\dd\Omega}{\dd x^j}\bom^i\right)=\frac{1}{\Omega} 
\left(\frac{\dd\Omega}{\dd x^i}\dd x^j-\frac{\dd\Omega}{\dd x^j}\dd
x^i\right)\,. 
\end{eqnarray} 
The Cartan curvature two-forms are then given by the second Cartan 
structure equation 
\begin{equation} 
{\Re^\mu}_\nu\equiv\dd{\bom^\mu}_\nu+{\bom^\mu}_\rho\wedge{\bom^\rho}_\nu \,, 
\end{equation} 
from which the components of the Riemann tensor can be read off via 
\begin{equation} 
{\Re^\mu}_\nu={\bkR^{\mu}}_{\nu\rho\s}\bom^\rho\wedge\bom^\s \,. 
\end{equation} 
Because of the symmetry of the Schwarzschild-de-Sitter spacetime, it 
is only necessary to evaluate the curvature 2-forms at a point, 
$x^1=\cdots=x^{(\dm-2)}=0$ say, giving 
\begin{equation} 
{\Re^t}_r=-\frac{{\cal V}''}{2}\bom^t\wedge\bom^r \,, \quad 
{\Re^t}_i=-\frac{{\cal V}'}{2r}\bom^t\wedge\bom^i \,, \quad 
{\Re^r}_i=-\frac{{\cal V}'}{2r}\bom^r\wedge\bom^i \,, \quad 
{\Re^i}_j=\left(\frac{k}{r^2}-\frac{{\cal V}}{r}\right)\bom^i\wedge\bom^j\,. 
\end{equation} 
Hence the non-zero components of the Riemann tensor are 
\begin{equation} 
{\bkR^{tr}}_{tr}=-\frac{{\cal V}''}{2} \,, \qquad 
{\bkR^{ti}}_{tj}={\bkR^{ri}}_{rj}=-\frac{{\cal V}'}{2r}{\delta^i}_j \,, \qquad 
{\bkR^{ij}}_{kl}=2\frac{k-{\cal V}}{r^2}{\delta^{[i}}_{[k} 
{\delta^{j]}}_{l]} \,, 
\end{equation} 
giving a the non-zero components of the Ricci tensor 
\begin{equation} 
{\bkR^t}_t={\bkR^r}_r=-\frac{{\cal V}''}{2}-(\dm-2)\frac{{\cal V}'}{2r}  
\,, \quad 
{\bkR^i}_j=\left((\dm-3)\frac{k-{\cal V}}{r^2}- 
\frac{{\cal V}'}{r}\right){\delta^i}_j \,, 
\end{equation} 
and the Ricci scalar 
\begin{equation} 
\bkR=-{\cal V}''-2(\dm-2)\frac{{\cal V}'}{r}+ 
(\dm-2)(\dm-3)\frac{k-{\cal V}}{r^2}\,. 
\end{equation} 
The Weyl tensor can then be calculated by the standard formula 
\begin{equation}\label{CR} 
{{\cal C}^{\al\be}}_{\gamma\delta}={\bkR^{\al\be}}_{\gamma\delta}- 
\frac{4}{\dm-2}{\delta^{[\al}}_{[\gamma}{\bkR^{\be]}}_{\delta]} 
+\frac{2}{(\dm-1)(\dm-2)}{\delta^{[\al}}_{[\gamma}{\delta^{\be]}}_{\delta]}\bkR \,, 
\end{equation} 
so that its non-zero components are 
\begin{equation} 
{{\cal C}^{tr}}_{tr}=(\dm-2){\cal W} \,, \qquad 
{{\cal C}^{ti}}_{tj}={{\cal C}^{ri}}_{rj}= 
-{\cal W}{\delta^i}_j \,, \qquad 
{{\cal C}^{ij}}_{kl}=\frac{4{\cal W}}{\dm-3} 
{\delta^{[i}}_{[k}{\delta^{j]}}_{l]} \, 
\end{equation} 
where the Weyl scalar $\cal W$ is obtained to be 
\begin{equation} 
{\cal W}=\frac{\dm-3}{(\dm-1)(\dm-2)}\left(-\frac{{\cal V}''}{2}+ 
\frac{{\cal V}'}{r}+\frac{k-{\cal V}}{r^2}\right) 
=(\dm-3)\frac{G_{_{[\dm]}}{\cal M}}{r^{\dm-1}} \, 
\end{equation} 
hence reducing to the expression (\ref{weylscalar}) on the brane.   
 
We can also determine the components of the normal vector $\norm^\mu$ in 
this coordinate system by contracting it with the timelike killing 
covector of the Schwarzschild-de-Sitter spacetime $k=dt$ and noting 
that $k_\mu\norm^\mu$ must be constant along geodesics and so depends only on 
$\tau$ in the brane-based coordinates~\cite{CU}.  With the normalization 
chosen in (\ref{GNansatz}), we have $k_\mu\norm^\mu=\dd a/\dd\tau$, 
which determines the $t$-component of $\norm^\mu$ in the static 
coordinates we have used to compute the Weyl tensor.  As $\norm^\mu$ is 
a unit vector and is orthogonal to the $(n-2)$-surfaces of spatial 
isometry, its components are 
\begin{equation} 
n^t=\frac{\dd a}{\dd\tau} \,, \qquad 
n^r={\cal V}^{1/2}\left(1+{\cal V}\left(\frac{\dd a}{\dd\tau}\right)^2\right)^{1/2} 
\,, \qquad n^i=0 \,. 
\end{equation} 
This allows us to calculate $\bulk_{\mu\nu}$ which, in the brane-based 
coordinates (\ref{GNansatz}), has the form given in (\ref{weylfield}).

\def\jnl#1#2#3#4#5#6{\hang{#1, {\it #4\/} {\bf #5}, #6 (#2).} } 
\def\jnltwo#1#2#3#4#5#6#7#8{\hang{#1, {\it #4\/} {\bf #5}, #6; {\it 
ibid} {\bf #7} #8 (#2).} }  
\def\prep#1#2#3#4{\hang{#1, #4.} } 
\def\proc#1#2#3#4#5#6{{#1 [#2], in {\it #4\/}, #5, eds.\ (#6).} } 
\def\book#1#2#3#4{\hang{#1, {\it #3\/} (#4, #2).} } 
\def\jnlerr#1#2#3#4#5#6#7#8{\hang{#1 [#2], {\it #4\/} {\bf #5}, #6. 
{Erratum:} {\it #4\/} {\bf #7}, #8.} }  
\def\prl{Phys.\ Rev.\ Lett.} 
\def\pr{Phys.\ Rev.}   
\def\pl{Phys.\ Lett.}   
\def\np{Nucl.\ Phys.} 
\def\prp{Phys.\ Rep.}   
\def\rmp{Rev.\ Mod.\ Phys.}   
\def\cmp{Comm.\ Math.\ Phys.}   
\def\mpl{Mod.\ Phys.\ Lett.}   
\def\apj{Astrophys.\ J.} 
\def\apjl{Ap.\ J.\ Lett.}   
\def\aap{Astron.\ Ap.}   
\def\cqg{Class.\ Quant.\ Grav.}   
\def\grg{Gen.\ Rel.\ Grav.}   
\def\mn{Mon.\ Not.\ Roy.\ Astro.\ Soc.} 
\def\ptp{Prog.\ Theor.\ Phys.}   
\def\jetp{Sov.\ Phys.\ JETP} 
\def\jetpl{JETP Lett.}   
\def\jmp{J.\ Math.\ Phys.}   
\def\zpc{Z.\ Phys.\ C}  
\def\cupress{Cambridge University Press}  
\def\pup{Princeton University Press}  
\def\wss{World Scientific, Singapore} 
\def\oup{Oxford University Press} 
\def\ijmp{Int.\ J.\ Mod.\ Phys.}

\end{document}